\documentclass[aps,pra,reprint,groupedaddress,amsmath,amssymb]{revtex4-1}
\usepackage{bm}
\usepackage{graphicx}
\usepackage[usenames, dvipsnames]{color}
\usepackage{subcaption}
\graphicspath{{figures/}}

\DeclareMathOperator{\tr}{Tr}

\usepackage{tikz}
\usepackage{ifthen}
\usepackage{tikz-3dplot}
\usepackage{pgfplots}
\usepackage{adjustbox}
\usepackage{xcolor}

\usepackage{amssymb}
\usepackage{amsmath}
\usepackage{physics}
\usepackage{graphicx}

\def\beq{\begin{eqnarray}}
\def\eeq{\end{eqnarray}}

\newcommand{\be}{\begin{equation}}
\newcommand{\ee}{\end{equation}}
\newcommand{\bea}{\begin{eqnarray}}
\newcommand{\eea}{\end{eqnarray}}

\renewcommand{\sb}{_\textsc{b}}

\renewcommand{\sb}{_\textsc{b}}

\begin{document}
\title{Fr\"ohlich-coupled qubits interacting with fermionic baths}%
\author{Erik Aurell}%
\email{eaurell@kth.se}
\affiliation{KTH – Royal Institute of Technology, AlbaNova University Center, 
SE-106 91 Stockholm, Sweden}%
\affiliation{Faculty  of  Physics,  Astronomy  and  Applied  Computer  Science,
Jagiellonian  University, 30-348  Krak\'ow,  Poland}
\author{Jan Tuziemski}%
\email{jan.tuziemski@fysik.su.se}
\altaffiliation[On leave from ]{Department of Applied Physics and Mathematics, Gdansk University of Technology}
\affiliation{Department of Physics, Stockholm University, AlbaNova University Center, Stockholm SE-106 91 Sweden}
\affiliation{Nordita, Royal Institute of Technology and Stockholm University,Roslagstullsbacken 23, SE-106 91 Stockholm, Sweden}

\date{\today}
\begin{abstract}
  We consider a quantum system such as a qubit, interacting with a bath
  of fermions as in the Fr\"ohlich polaron model. The interaction Hamiltonian is thus
  linear in the system variable, and quadratic in the fermions.
  Using the recently developed extension of Feynman-Vernon theory to non-harmonic baths
  we evaluate quadratic and the quartic terms in the influence action. We find that for this
  model the quartic term vanish by symmetry arguments.
  Although the influence of the bath on the system is of the same form as 
  from bosonic harmonic oscillators up to effects to sixth order
  in the system-bath interaction, the temperature dependence is nevertheless
  rather different, unless rather contrived models are considered.
\end{abstract}

\maketitle

\section{Introduction}
\label{sec:intro}
The theory of open quantum systems 
has attracted increased attention
in recent years, 
motivated by advances quantum information theory~\cite{Wilde-book}
and emerging quantum technologies~\cite{DevoretWallrafMartinis,WendinShumeiko}.
For these to become practically useful in a broad range of applications
a main roadblock to overcome is the strong tendency of large quantum systems 
to turn classical due to interactions with the
rest of the world~\cite{HarocheRaimond1996,Zurek2003,Schlosshauer-book}.
Open quantum systems encompass the various concepts and analytic and numerical techniques that have been
developed to describe and estimate the development of a quantum system interacting
with an environment~\cite{Weiss-book,open}.

A special place in open quantum system theory belongs to 
problems where a general system (the system of interest)
interacts linearly with one or several baths of harmonic oscillators.
One reason is that 
resistive elements
in a small electrical circuit
can be modeled  as many LC elements in parallel, of which each one 
obeys the equation of a harmonic oscillator.
At very low temperature as in quantum technology applications,
these harmanic oscillators should be quantised~\cite{Devoret1995}.
A related reason is the number of physical environments (phonons, photons)
that can also be directly described this way.
In the Lagrangian formulation of quantum mechanics~\cite{FeynmanHibbs} 
the development of a wave function (unitary operator $U$) is described by a path integral,
while the development of a density matrix 
(quantum operation $U\cdot U^{\dagger}$) is described by 
two path integrals, one (forward path) for $U$ and one (backward path) for $U^{\dagger}$.
A third reason why harmonic oscillator baths are interesting
is that the paths of such baths can be integrated out yielding the famous 
Feynman-Vernon theory~\cite{Feynman1963}.
The only trace of the bath (or baths) is then the Feynman-Vernon action,
quadratic terms in the forward and backward paths.

Nevertheless, most physical environments do only approximately
or not at all consist of degrees of freedom that can be described
as bosonic harmonic oscillators.
Conduction band electrons in normal metals
are for instance obviously fermions. Even if these fermions 
by themselves are free (and hence can be treated as fermionic harmonic oscillators),
in the open quantum system context it is their interaction with the system of interest that counts.
If that system is a quantum variable 
such as a qubit, the simplest interaction
that can be considered is quadratic in the fermionic
variables and linear in the system of interest.
As a term in an interaction Hamiltonain
that is $X a b$
where  $X$ is the quantum variable of the system of interest
and $a$ and $b$ are creation or destruction operators of
the fermions.
Interaction Hamiltonians of this type appear
in the Fr\"ohlich polaron model of 
the motion of a
conduction electron in an ionic crystal~\cite{Froelich54,DevreeseAleksandrov2009}.
In Feynman's variational treatment, one electron is 
modelled as a non-relativistic particle interacting with a bath of bosonic harmonic operators 
which are then integrated out.
Here we are interested in the opposite case where one bosonic degree of freedom,
\textit{i.e.} the qubit, 
describes the system of interest,
and we want to ``integrate out'' the fermions.
One problem with such an approach is that fermionic functional integrals (Grassman integrals)
are mathematically non-trivial objects. Another is that for the Fr\"ohlich-like coupling
both the bath Hamiltonian and the interaction are quadratic in the fermionic degrees
of freedom; the result is hence two fermionic functional determinants depending
on the forward and backward histories of the system of interest acting as external fields.

An approach to similar problems, used for a long time in condensed matter theory,
is Keldysh techniques~\cite{KELDYSH1,KELDYSH2}. While essentially equivalent to
Feynman-Vernon theory, Keldysh theory was developed for other applications, and encompassing from
the start fermionic baths. The kernels of the quadratic terms in Feynman-Vernon theory
can thus be identified with pair-wise bath correlation functions, in Keldysh theory
referred to as ``dressed non-equilibrium Greens functions''.
Here we will instead follow the recently developed 
extension of the Feynman-Vernon theory to
non-harmonic baths~\cite{AurellGoyalKawai2019}.
One advantage of this approach is that it gives access
also to terms in Feynman-Vernon influence functional higher than
quadratic. Let us remark that from the functional integral point
of view it is obvious that such terms must exist: while the
bath can always be integrated out in principle, it is
only for harmonic (bosonic or fermionic) baths that all the integrals are Gaussian
and can be done in closed form.
A main result of~\cite{AurellGoyalKawai2019} is that higher-order Feynman-Vernon terms
depend on cumulants of bath correlation functions.
The first non-standard term in the extended Feynman-Vernon theory for the
dynamics of the system hence involves fourth-order cumulants of the correlation functions
of the compound bath variables $a b$ \textit{i.e.} eighth-order fermionic correlations.
Perhaps suprisingly we find that for the Fr\"ohlich-coupled system these terms actually cancel in
the influence function. 
\\\\
The paper is organized as follows.
In Section~\ref{sec:problem}
we state the problem and make general remarks
of what one can expect of the solution.
In \ref{sec:spin-boson} we assume as a concrete
example that the variable is a qubit (a two-state system)
coupled to the bath as in the spin-boson problem,
and state more precisely the system-bath interaction we study in the rest of the paper. 
In Section~\ref{sec:presentation} we present the
structure of the first term of the Feynamn-Vernon action and state that the second term in the expansion of the action vanishes in our model. Here we also sketch calculations of  bath correlation functions of interest in our theory.
In Section~\ref{sec:analysis} the standard (second order) Feynamn-Vernon action of the considered model is compared to that of a harmonic bosonic bath. 
Appendices \ref{sec:non-harmonic}, \ref{sec:generalized-FV} contain summaries of technical details from~\cite{AurellGoyalKawai2019}, included for completeness. Appendix \ref{app:4th} presents the detailed argument that in the model considered the fourth order cumulant vanishes, and therefore there is no fourth order contribution to the generalized Feynman-Vernon action.   

\section{Statement of the problem}
\label{sec:problem}
Let us consider a system consisting of one bosonic variable
and a bath of free fermions as discussed above. That means a Hamiltonian
\begin{equation}
\label{eq:bath-free}
\hat{H}_{TOT} = \hat{H}_{S} + \hat{H}_{INT} +  \hat{H}_{B},
\end{equation}
where the first term $\hat{H}_{S}$ is the Hamiltonian of the system.
For a bosonic variable the evolution operator corresponding
to $\hat{H}_{S}$ can be written as a path integral
\begin{equation}
U_S=e^{-\frac{i}{\hbar}H_S \tau} = \int {\cal D}X e^{\frac{i}{\hbar}S[X]},
\end{equation}
where $S[X]$ is the action of path $X$. The evolution operator acting on density matrices
is similarly a double path integral over a ``forward path'' and a ``backward path''
\begin{equation}
\label{eq:quantum-map-pure}
U_S\cdot U_S^{\dagger} = \int {\cal D}X  {\cal D}Y e^{\frac{i}{\hbar}S[X]-\frac{i}{\hbar}S[Y]} \cdot
\end{equation}
where the slot marks where the initial density matrix is to be inserted.

The free bath Hamiltonian in \eqref{eq:bath-free} is 
\begin{equation}
\label{eq:H_B}
\hat{H}_{B} = \sum_k \epsilon_k \hat{c}^{\dagger}_k \hat{c}_k,
\end{equation}
where $\hat{c}^{\dagger}_k$ ($\hat{c}_k$)
is the creation (destruction) of fermions, and the interaction Hamiltonian is of the type (below we will use a more
specific model)
\begin{equation}
\hat{H}_{INT} = X \sum_{k,l} g_{kl} \hat{c}^{\dagger}_k \hat{c}_l.
\end{equation}

Initially the bath and the system are assumed
independent, and the bath is in thermal equilibrium at inverse temperature $\beta$.
The evolution operator of the system is the quantum map
(or quantum operator) given by
\begin{equation}
\label{eq:quantum-map-dissipative}
\Phi \cdot = \hbox{Tr}_B\left[U\left(\rho_B(\beta)\otimes \cdot\right)U^{\dagger}\right]
\end{equation}
where $\hat{U}=e^{-\frac{i}{\hbar}(\hat{H}_S+\hat{H}_{INT}+\hat{H}_B) \tau}$ is the total evolution operator of the combined
system and bath, $\rho_B(\beta)$ is the initial equilibrium density matrix of the bath,
and $\cdot$ marks where to insert the initial density matrix of the system.

Suppose that the evolution of the bath can also be written as a double
path integral. If so the bath can be integrated out, so that we have
\begin{equation}
\label{eq:quantum-map-dissipative-2}
\Phi \cdot = \int {\cal D}X  {\cal D}Y e^{\frac{i}{\hbar}S[X]-\frac{i}{\hbar}S[Y]} {\cal F}[X,Y] \cdot
\end{equation}
The new term compared to \eqref{eq:quantum-map-pure} is the Feynman-Vernon influence
functional, \textit{i.e.} what remains after integrating out
the bath paths while the system paths are held fixed.
Although important general properties of the influence functional were
stated in~\cite{Feynman1963}, in practice this formalism has mostly been used for when the baths are free bosons interacting 
linearly with a system. In that case all the path integrals over the baths
are Gaussians, and ${\cal F}$ can be written as $e^{\frac{i}{\hbar}S_i[X,Y]-\frac{1}{\hbar}S_r[X,Y]}$
where $S_i[X,Y]$ and $S_r[X,Y]$ are two explicit quadratic functionals 
of the forward and backward system paths, usually known as Feynman-Vernon action.

On the other hand, it is not necessary to assume that the bath can be represented as path
integrals. As reviewed in~\cite{open} and rederived in \cite{AurellGoyalKawai2019},
the super-operator $\Phi$ in \eqref{eq:quantum-map-dissipative} can be computed 
perturbatively, and the terms translated back to a double path integral
over the system. In this way one can identify the kernels 
in the actions $S_i[X,Y]$ and $S_r[X,Y]$ as being equilibrium pair correlations
in the bath. Importantly this holds for any equilibrium bath.
The price to pay if the bath is not harmonic is that there are higher-order
terms that are respectively fourth, sixth etc order in the system variables $X$ and $Y$.

\section{A qubit coupled to a fermionic bath as in spin-boson problem}
\label{sec:spin-boson}
For concreteness, and since this would be a main application to quantum information science,
we now assume that the system of interest is a
a qubit (a two-state system) governed by a system Hamiltonian
\begin{equation}
\label{eq:H_S}
\hat{H}_{S} = \frac{\varepsilon}{2}\hat{\sigma}_z + \hbar\frac{\Delta}{2}\hat{\sigma}_x .
\end{equation}
The evolution operator $e^{-\frac{i}{\hbar}H_S \tau}$
can be represented by inserting resolution of the identity between very
small time increments $\delta\tau$. The first term in 
\eqref{eq:H_S} then only contributes if the state stays the
same between two small time increments; that contribution is
$e^{\pm \frac{i}{\hbar}\frac{\varepsilon}{2} \delta\tau}$. The parameter
$\varepsilon$ is hence the \textit{level splitting}.
The second term in \eqref{eq:H_S} on the other hand
only contributes if the state changes over a small time increment, and the contribution is
$(\pm i\frac{\Delta}{2} \delta\tau)$. The parameter
$\Delta/2$, which has dimension of a rate, is hence the \textit{tunelling element}.  

The paths in $X$ and $Y$ in the path integral in \eqref{eq:quantum-map-pure}
are nothing but a way to represent $e^{-\frac{i}{\hbar}\hat{H}_S \tau}$ and $e^{\frac{i}{\hbar}\hat{H}_S \tau}$,
and are hence \textit{piece-wise constant}, equal to $\pm 1$.
Before continuing we note a clash of conventions: $X$ and $Y$ are
in the literature on open quantum systems used to refer to the history of a system variable
which is intergrated over.
In our case these are the histories (forward and backward)
of a representation of $\hat{\sigma}_z$, and $\hat{X}$ is also used for the system part of the interaction Hamiltonian. 
This is the convention we follow.
In the quantum information
literature $\hat{X}$ and $\hat{Y}$ instead refer to the operators $\hat{\sigma}_x$
and $\hat{\sigma}_y$, while the operator $\hat{\sigma}_z$ is written $\hat{Z}$.
We do not follow this convention. 

Now, it is convenient to include the contributions from the level splitting in the
actions in \eqref{eq:quantum-map-pure}, and the contributions from the tunelling
elements in the path measures ${\cal D}X$ and ${\cal D}Y$.
If so  ${\cal D}X$ and ${\cal D}Y$ are nothing but the path probabilities 
of (classical) Poisson point processes, except that the jump rates are
purely imaginary. That is, we can interpret $X$ as $s_i,n,t_1,\ldots,t_n$
where $s_i$ is the initial state (up or down), $n$ is the number of jumps
and $t_1<t_2<\ldots$ are the jump times. The purely imaginary path measures are then
\begin{equation}
\label{eq:DX}
\int {\cal D}X \left(\cdot\right) = \sum_n \prod_{s=1}^n (\pm i\frac{\Delta}{2}) \int dt_s \left(\cdot\right).
\end{equation}
The advantage of the above is that it can accomodate also a coupling to a bath when that
coupling is proportional to $\sigma_z$.
When the bath is composed of bosonic harmonic oscillators this is
the \textit{spin-boson problem}; the above path integral was developed
by Leggett and collaborators for that problem in \cite{Leggett87}.

For our problem we will consider
the interaction Hamiltonian is
\begin{eqnarray}
\label{eq:H-INT-Schrodinger-rep}
\hat{H}_{INT} &=& \sum_{\substack{k,l}}g_{kl}\hat{X}\left( \hat{c}_k + \hat{c}_k^{\dagger} \right) \left( \hat{c}_l  + \hat{c}_l^{\dagger}\right), 
\end{eqnarray}
where $\hat{X}$ ($\hat{\sigma}_z$) is the system part of the interaction, 
$\hat{c}^{\dagger}_l,\hat{c}^{\dagger}_k$ ($\hat{c}_k,\hat{c}_l$)
are the creation (destruction) operators of two fermions,
and $g_{kl}$ is a coupling constant. Due to the anti-commutation rules for fermions we can set $g_{kl} = -g_{lk}$.

For the following sections it is convenient to introduce 
an interaction representation based on  
\eqref{eq:H_B} and \eqref{eq:H_S}.
Bath destcruction operators transform as
\begin{equation} 
\hat{c}_i(t) = e^{\frac{i}{\hbar}\hat{H}_B t} \hat{c}_i   e^{-\frac{i}{\hbar}\hat{H}_B t} = \hat{c}_i e^{-i\omega_i t},
\end{equation}
where $\omega_i \equiv \epsilon_i/\hbar$, and bath creation operators as $\hat{c}^{\dagger}_i(t) = \hat{c}^{\dagger}_i e^{i\omega_i t}$. 
Explicit form of the transformed system operator $\hat{X}$ in \eqref{eq:H-INT-Schrodinger-rep} is not relevant for further considerations. 
In this representation the interaction Hamiltonian is
\begin{eqnarray}
\label{eq:H-INT-interaction-rep}
\hat{H}_{INT}(t) &=& \sum_{k,l} g_{kl}X(t)\left( \hat{c}_k e^{-i\omega_k t}+ \hat{c}_k^\dagger e^{i\omega_kt} \right)\nonumber \\
&& \left( \hat{c}_l e^{-i\omega_l t}+ \hat{c}_l^\dagger e^{i\omega_l t}\right).
\end{eqnarray}
We also assume that the bath is initially in a thermal state where $\rho = e^{-\beta \sum_k E_k \hat{c}^\dagger_k \hat{c}_k}$ with $\beta \equiv 1/(k_B T)$.  

\section{The generalized Feynman-Vernon action terms}
\label{sec:presentation}
 In most cases  discussed in literature Feynman-Vernon action is of the second order in the  system paths. This occurs e.g. for a system interacting linearly with a bath of free bosons. However, for other type of baths and couplings higher order terms in the action appear. A systematic way of dealing with such situations was formulated in \cite{AurellGoyalKawai2019} and for the convenience of the reader is summarized in Appendix \ref{sec:non-harmonic} and \ref{sec:generalized-FV}. In that approach the total Feynman-Vernon action is expressed as a sum of different-order terms (i.e. involving different number of system paths). In the Appendix \ref{sec:generalized-FV} we show that expression for the the usual quadratic Feynman-Vernon action can be rewritten such that    
\begin{eqnarray}
S^{(2)}=-\frac{\hbar}{2}\int_{t_0}^{t}\dd{t_1}  \int_{t_0}^{t}\dd{t_2} C(t_1,t_2) \mathcal{J}\left[Y_{t_2},Y_{t_1},X_{t_1},X_{t_2}\right],   \nonumber
\label{eq:quadraticFV}
\end{eqnarray}
 where $\mathcal{J}\left[\ldots\right]$ is a quadratic functional over paths of the system, which explicit form is given by Eq. (\ref{eq:Adquadratic}), $X_s$ and $Y_s$ are forward and backward path of the system evaluated at time $s$, and $C(t_1,t_2)$ is  the bath correlation function. For the problem considered here it reads
 \begin{eqnarray}
 \label{eq:2times}
C(t_1,t_2) &=& \\ &&-\sum_{k,l} g^2_{kl}\langle \hat{Q}_k(t_1) \hat{Q}_k(t_2) \rangle \langle \hat{Q}_l(t_1) \hat{Q}_l(t_2) \nonumber \rangle,   
 \end{eqnarray}
 where $\hat{Q}_k(t_i) \equiv \hat{c}_k e^{-i\omega_k t_i}+ \hat{c}_k^\dagger e^{i\omega_k t_i} $ and
 \begin{eqnarray}
 &&\left \langle \hat{Q}_k(t_1) \hat{Q}_k(t_2) \right \rangle \equiv  \\ \nonumber &&\cos \omega_k \left(t_1-t_2\right)  -i \sin\omega_k \left(t_1-t_2\right)  \tanh {\frac{\beta E_k}{2}},
 \end{eqnarray}
 is thermal expectation value of fermionic operators.
 Derivation of the above result relies on two simple facts. The first is that, in general, consecutive action terms depend on the following bath correlation functions
 \begin{equation}
 \begin{aligned}
& C(t_1,\ldots,t_{2n}) = \sum_{k_1, \ldots k_{2n}, l_1, \ldots, k_{2n}} g_{k_1 l_1} \ldots g_{k_{2n} l_{2n}} \times\\ \nonumber &Tr(\hat{Q}_{k_1}(t_1)\hat{Q}_{l_1}(t_1) \dots \hat{Q}_{k_{2n}}(t_{2n})\hat{Q}_{l_2n}(t_{2n}) \rho ).
\nonumber
\end{aligned}
\end{equation}
The above is non-zero only if the number of all fermion indices $k$'s and $l$'s is even. For the quadratic action term we find that the only non-zero contribution is    
  \begin{equation}
 \begin{aligned}
 &Tr(\hat{Q}_{k}(t_1)\hat{Q}_{l}(t_1)\hat{Q}_{k}(t_2)\hat{Q}_{l}(t_2)\rho ) = 
 \nonumber  \\
 &-Tr(\hat{Q}_k(t_1)\hat{Q}_k(t_2) \hat{Q}_l(t_1)\hat{Q}_l(t_2)   \rho ),  \nonumber
 \end{aligned}
 \end{equation}
 from which Eq. (\ref{eq:2times}) immediately follows.

  The third  order term is automatically zero as it contains odd number of indices $k$ and $l$.

 The fourth order term reads
 \begin{eqnarray}
 \label{eq:operatorFV4thMT}
 &&S^{(4)}=\frac{(-i)^{4}}{4 !\hbar^4} \int_{t_{i}}^{t_{1}} \mathrm{d} t_{1} \int_{t_{i}}^{t_{f}} \mathrm{d} t_{2} \int_{t_{i}}^{t_{f}} \mathrm{d} t_{3}
 \int_{t_{i}}^{t_{f}} \mathrm{d} t_{4}
 \sum_{d_{1}, d_{2}, d_{3},d_{4}} \\ \nonumber &&G_{4}^{d_{1}, d_{2} ; d_{3}, d_{4}}\left(t_{1}, t_{2}, t_{3}, t_4\right) \mathcal{X}_{\mathrm{s}}^{d_{1}}\left(t_{1}\right) \mathcal{X}_{\mathrm{s}}^{d_{2}}\left(t_{2}\right) \mathcal{X}_{\mathrm{s}}^{d_{3}}\left(t_{3}\right) \mathcal{X}_{\mathrm{s}}^{d_{4}}\left(t_{4}\right),
 \end{eqnarray}
 where
 \begin{equation}
\label{eq:4thcummulant}
\begin{aligned}
&G_{4}^{d_{1}, d_{2}, d_{3}, d_{4}}\left(t_{1}, t_{2}, t_{3}, t_{4}\right) =\\&C^{d_{1}, d_{2}, d_{3}, d_{4}}\left(t_{1}, t_{2}, t_{3}, d_{4}\right)-C^{d_{1}, d_{2}}\left(t_{1}, t_{2}\right) C^{d_{3}, d_{4}}\left(t_{3}, t_{4}\right) \\
&-C^{d_{1}, d_{3}}\left(t_{1}, t_{3}\right) C^{d_{2}, d_{4}}\left(t_{2}, t_{4}\right)-C^{d_{1}, d_{4}}\left(t_{1}, t_{4}\right) C^{d_{2}, d_{3}}\left(t_{2}, t_{3}\right),
\end{aligned}
\end{equation}  
 is the fourth order super-operator cumulant involving super-operator correlation functions of the bath
 \begin{eqnarray}
 \begin{aligned}
 &C^{d_{1}, \cdots, d_{2n}}\left(t_{1}, \cdots, t_{n}\right) = \\& \sum_{k_1, \ldots k_{2n}, l_1, \ldots, k_{2n}} g_{k_1 l_1} \ldots g_{k_{2n} l_{2n}} \times \\  &Tr\left[\overleftarrow{\mathcal{T}} (\hat{Q}^{d_1}_{k_1}(t_1)\hat{Q}^{d_2}_{l_1}(t_1) \dots \hat{Q}^{d_{2n-1}}_{k_{2n}}(t_{2n})\hat{Q}^{d_{2n}}_{l_2n}(t_{2n}) ) \rho\right].
 \end{aligned}
 \end{eqnarray}
 Indices $d_{i}+\pm$ indicate, on which side of the environment density matrix an operator acts (left or right for $+$, $-$ respectively). As for now $d_{1}, \cdots d_{4}$ will be fixed and dependence on them will be dropped. We show that, due to cancellations, there is no fourth order contribution to the Feynman-Vernon action. Here we present the main steps of the argument, the details can be found in  Appendix \ref{app:4th}.  First of all one needs to calculate the fourth order correlation function, which for our model reads
 \begin{eqnarray}
 \label{eq:4timecorfunMT}
 C(t_1,t_2,t_3,t_4) = \sum_{k_1,l_1, \ldots k_4,l_4}g_{k_1l_1} \ldots g_{k_4l_4} \times \\ Tr \left[\hat{Q}_{k_1}(t_1)\hat{Q}_{l_1}(t_1)\ldots \hat{Q}_{k_4}(t_4)\hat{Q}_{l_4}(t_4) \rho \right]. \nonumber \\
 \end{eqnarray}
Note  that integration in Eq. (\ref{eq:operatorFV4thMT}) is performed with respect to un-ordered times. To avoid confusion, the time-ordered times will be referred to as $s_i$. The non-zero terms in the sum Eq. (\ref{eq:4timecorfunMT}), are those in which a given index e.g. $k_i$ appears an even number of times. Therefore we can distinguish the following cases:\\
1. Pair-wise groupings. Here a given index appears only twice. In the Appendix  \ref{app:4th} we show that it is sufficient to consider forming pairs out of $k_i$ and $l_j$ indices respectively. An example of such a grouping is  $k_1=k_2=k,\; k_3=k_4=k'$ and a similar pairing for $l_j$. This term reads
\begin{eqnarray}
g^2_{kl}g^2_{k'l'}Tr&&\left[\hat{Q}_{k}(s_1)\hat{Q}_{k}(s_2)\hat{Q}_{k'}(s_3)\hat{Q}_{k'}(s_4)\right. \\ \nonumber &&\times \left. [\hat{Q}_{l}(s_1)\hat{Q}_{l}(s_2)\hat{Q}_{l'}(s_3)\hat{Q}_{l'}(s_4) \right] = \\
g^2_{kl}g^2_{k'l'}Tr&&\left[ [12]_{k}[34]_{k'} [12]_{l}[34]_{l'} \right],
\end{eqnarray}  
where the expression was written using time ordered times $s_i$ and we introduced new notation $[ij]_k \equiv \hat{Q}_{k}(s_i)\hat{Q}_{k}(s_j)$, which will be helpful in further considerations. If the super-operator indices   $d_{1}, \cdots d_{4}$ are fixed we can relate time-ordered times to un-constraint times  as $s_1=t_x,\;, s_2=t_y, \; s_3=t_z \; ,s_4=t_w$. Now we consider a new time ordering (depending on $d_1, \ldots d_4$) of operators $\hat{Q}_{m}(s_i)$, where $m \in \{k,k,l,l' \}$
\begin{eqnarray}
S_{d,m}(a,b) = T_{d,m}(x,y) = &&\overrightarrow{\mathcal{T}} \prod_{i\in (x,y); d_i=-} Q_m(t_i) \times \\ &&\overleftarrow{\mathcal{T}}\prod_{j\in (x,y); d_i=+} Q_m(t_j),
\end{eqnarray}
if the order of the operators is the same as $(x, y)$, and
\begin{eqnarray}
S_{d,m}(a,b) = -T_{d,m}(x,y),
\end{eqnarray}
for the opposite order. In this way we can rewrite the term $[12]_{k}[34]_{k'}$ as $S(1234 \rightarrow xyzw)S_{d,k}(x,y)S_{d,k}(w,z)$, where $S(p \rightarrow q)$ is a permutation sign. Subsequently, in the Appendix  \ref{app:4th} we show that pair-wise groupings terms can be rewritten with as sum over all such permutations. We examine properties of those permutations under change of time variables and exchange of indices and show which terms cancel. The final contribution from the pair-wise groupings is found to be 
\begin{eqnarray}
\label{eq:counterIresMT}
\sum_{k,l,k',l'}g_{kl}^2 g_{k'l'}^2  &&\left( \left \langle 12 \right \rangle_{k} \left \langle 34 \right \rangle_{k'}   \left \langle 12 \right \rangle_{l} \left \langle 34 \right \rangle_{l'} \right. +  \nonumber \\ && \left \langle 13 \right \rangle_{k} \left \langle 24 \right \rangle_{k'}   \left \langle 13 \right \rangle_{l} \left \langle 24 \right \rangle_{l'} + \nonumber \\ && \left. \left \langle 14 \right \rangle_{k} \left \langle 23 \right \rangle_{k'}   \left \langle 14 \right \rangle_{l} \left \langle 23 \right \rangle_{l'} \right),
\end{eqnarray}   
where $\left \langle ab \right \rangle_{m}\equiv \left \langle \hat{Q}_m(t_a) \hat{Q}_m(t_b) \right \rangle $.  More detailed discussion can be found in Appendix \ref{sub:subI}. \\
2. Four $k$ grouping, pair-wise $l$ grouping. An example of such a term (in the operator notation) is 
\begin{equation}
\sum_{k,l,l'}g_{kl}^2 g_{kl'}^2 Tr \left[[1234]_k \left([12]_{l}[34]_{l'}-[13]_{l}[24]_{l'}+[14]_{l}[23]_{l'}\right)\right].
\end{equation}   
 The $k-$th part of this expression can be evaluated using Wicks theorem. Then one applies essentially the same arguments as those mentioned in the previous case, as the reasoning does not rely on the particular arrangement of $k,l$ indices. Thus the final contribution of those terms is found to be
 \begin{eqnarray}
 \label{eq:caseIIresMT1}
 \sum_{k,l,l'}g_{kl}^2 g_{kl'}^2&&\left(\left \langle 12 \right \rangle_{k} \left \langle 34 \right \rangle_{k} \left \langle 12 \right \rangle_{l} \left \langle 34 \right \rangle_{l'} \right.  + \nonumber \\ &&\left \langle 13 \right \rangle_{k} \left \langle 24 \right \rangle_{k} \left \langle 13 \right \rangle_{l} \left \langle 24 \right \rangle_{l'} +\nonumber \\ &&\left. \left \langle 14 \right \rangle_{k} \left \langle 23 \right \rangle_{k} \left \langle 14 \right \rangle_{l} \left \langle 23 \right \rangle_{l'} \right).
 \end{eqnarray}  \\
 3. Four $k$,$l$ grouping. In the operator notation we can write
 \begin{eqnarray}
 \sum_{kl}g^4_{kl}Tr \left[[1234]_k [1234]_l\right],
 \end{eqnarray}
 We use Wicks theorem and arguments from previous cases to show that these trms equal
 \begin{eqnarray}
 \label{eq:caseIIIresMT1}
 \sum_{k,l}g_{kl}^4&&\left(\left \langle 12 \right \rangle_{k} \left \langle 34 \right \rangle_{k} \left \langle 12 \right \rangle_{l} \left \langle 34 \right \rangle_{l} \right.  + \nonumber \\  &&\left \langle 13 \right \rangle_{k} \left \langle 24 \right \rangle_{k} \left \langle 13 \right \rangle_{l} \left \langle 24 \right \rangle_{l} + \nonumber \\ &&\left. \left \langle 14 \right \rangle_{k} \left \langle 23 \right \rangle_{k} \left \langle 14 \right \rangle_{l} \left \langle 23 \right \rangle_{l} \right),
 \end{eqnarray}
Finally we need to subtract from above results the counter-terms from Eq. (\ref{eq:4thcummulant}) i.e. the products of two-times correlation functions. Direct calculation shows (see Appendix \ref{app:4th}) that there is a total cancellation of those terms. As a result, for the considered model there is no contribution from the fourth order term.

\section{Physical analysis of the quadratic action terms}
\label{sec:analysis}
In this section we analyze the quadratic term of the action. Our aim is to compare it to the action for a harmonic bosonic bath that is linear coupled to a system. The easiest way of doing this is by rewriting the Feynman-Vernon action with the help of imaginary and real parts of the kernel $k^I(t)$ and $k^R(t)$ respectively. The general expression reads
\begin{eqnarray}
S^{(2)} =-\frac{\hbar}{2}\int_{t_f}^{t}\dd{t}  \int_{t}^{t}\dd{s} &&\left(X_t-Y_t \right) k_{R}(t-s)\left(X_s+Y_s \right) +\\ &&\left(X_t-Y_t \right) k_{I}(t-s)\left(X_s-Y_s \right),
\end{eqnarray}  
where $X_t,Y_t$ correspond to forward and backward path of a system operator. For the fermionic bath considered here the kernels $k^I(t)$ and $k^R(t)$ are respectively:  
\begin{equation}
\begin{aligned}
&k^I_F(t_1-t_2)= \\ \nonumber &2i \sum_{k,l} g^2_{kl} \left[\sin \left[\omega_k \left( t_1-t_2\right)\right]  \cos \left[\omega_l \left( t_1-t_2\right)\right] \tanh \frac{\beta E_k}{2} \right. \\ & \left.+\cos \left[ \omega_k \left( t_1-t_2\right)\right] \sin \left[ \omega_l \left( t_1-t_2\right)\right]  \tanh \frac{\beta E_l}{2} \right] \\
&k^R_F(t_1-t_2)= \\ \nonumber -&2 \sum_{k,l} g^2_{kl} \left[\cos \left[ \omega_k \left( t_1-t_2\right)\right] \cos \left[ \omega_l \left( t_1-t_2\right)\right]  \right. \\ & \left.-\sin \left[ \omega_k \left( t_1-t_2\right)\right] \sin \left[\omega_l \left( t_1-t_2\right)\right] \tanh \frac{\beta E_k}{2}  \tanh \frac{\beta E_l}{2} \right],
\end{aligned}
\end{equation}
whereas for bosonic baths coupled linearly to the system (see e.g. \cite{open}): 
\begin{equation}
\begin{aligned}
&k^I_{B}(t_1-t_2)=i\sum_{k} \frac{c^2_{k}}{2 m_{k} \omega_{k}} \sin \omega_{k}\left(t_1-t_2\right)\\
&k^R_{B}(t_1-t_2)=\sum_{k} \frac{c^2_{k} }{2 m_{k} \omega_{k}} \operatorname{coth}\left(\frac{\beta \omega_{k} }{2}\right) \cos \omega_{k}\left(t_1-t_2\right).
\end{aligned}
\end{equation}
 Let us now discuss differences and similarities between those expressions. Imaginary kernels modify action and hence describe dissipation. For harmonic bosonic baths imaginary kernel is temperature independent, what is not the case for the model consider here. However, we can consider two temperature regimes with a simpler behavior.
 In the low temperature regime  ($\beta \gg 1$) the fermionic kernel resembles the bosonic one  $k^I(s-u)\approx  2 i  \sum_{k,l} g^2_{kl} \sin\left[\left(\omega_k+\omega_l\right)\left(s-u\right)\right]$, with frequency of a bosonic mode replaced by sum of frequencies of interacting fermions. In the opposite regime i.e. high temperatures ($\beta \ll  1$) the dissipation kernel vanishes. On the other hand, the real kernel introduces noise and is responsible for decoherence process. In the bosonic case decoherence strength increases with temperature. For the fermionic model in the low temperature limit  ($\beta \gg 1$) the real kernel   $k^R(s-u)\approx -2 \sum_{k,l} g^2_{kl} \cos\left[\left(\omega_k+\omega_l\right)\left(s-u\right)\right] $ is similar to the bosonic one. However, magnitude of the fermionic kernel does not grow with temperature: The high temperature limit ($\beta \ll 1$)  of the real kernel reads     $k^I(s-u)\approx -2\sum_{k,l} g^2_{kl} \cos \left[ \omega_k \left( s-u\right)\right] \cos \left[ \omega_l \left( s-u\right) \right] $. As we can see, for the low temperatures the fermionic bath behaves similarly to the bosonic one and the differences between them are most important in the high temperature regime.  

The formalism described here is general and can be applied also to bosonic systems coupled aquatically to a free bosonic bath. In such a case fermionic operators $\hat{c}_k,\hat{c}^{\dagger}_k$ are replaced with their bosonic counterparts $\hat{a}_k,\hat{a}^{\dagger}_k$ that obey the canonical commutation relations $ \left[\hat{a}_k,\hat{a}^{\dagger}_l \right] =\delta_{k,l}$ (other commutators vanish). Apart from this change the form of the interaction Hamiltonian Eq. (\ref{eq:quadraticFV}) and the free bath Hamiltonian Eq. (\ref{eq:H_B}) remains the same. Therefore, the structure of the results is similar to the discussed fermionic case, and the differences steam from the different (commutation) relations for the bosonic operators. In particular we find that the second order action for the bi-linear coupling to the bosonic bath is
\begin{equation}
\begin{aligned}
&k^I_{BB}(t_1-t_2)= \\ \nonumber &2i \sum_{k,l} g^2_{kl} \left[\sin \left[\omega_k \left( t_1-t_2\right)\right]  \cos \left[\omega_l \left( t_1-t_2\right)\right] \coth \frac{\beta E_k}{2} \right. \\ & \left.+\cos \left[ \omega_k \left( t_1-t_2\right)\right] \sin \left[ \omega_l \left( t_1-t_2\right)\right]  \coth \frac{\beta E_l}{2} \right] \\
&k^R_{BB}(t_1-t_2)= \\ \nonumber &2 \sum_{k,l} g^2_{kl} \left[\cos \left[ \omega_k \left( t_1-t_2\right)\right] \cos \left[ \omega_l \left( t_1-t_2\right)\right] \times  \right. \\ & \left. \coth \frac{\beta E_k}{2} \coth \frac{\beta E_l}{2} -\sin \left[ \omega_k \left( t_1-t_2\right)\right] \sin \left[\omega_l \left( t_1-t_2\right)\right]    \right].
\end{aligned}
\end{equation}
From the above one can see that the following substitution allows to recover results for the bosonic bi-linear bath from the fermionic one  
\begin{eqnarray}
&&k^I_{BB}(t_1-t_2) =  \coth \frac{\beta E_k}{2}  \coth \frac{\beta E_l}{2} k^R_{F}(t_1-t_2),
\end{eqnarray}
 and the same relation holds for the real part of the kernel $k^R_{BB}(t_1-t_2)$.     
\section{Discussion}
\label{sec:discussion}
In this paper we addressed the model of a quantum variable such as a qubit interacting with a fermionic bath. The coupling between the qubit  and the bath is quadratic in fermionic operators, and the bath is initially in a thermal state. To investigate this system we employed the extension of the Feynman-Vernon influence functional technique that allows to systematically study higher order contributions to the Feynamn-Vernon action that arise from system-bath interaction being non-linear with respect to bath operators. We explicitly computed the second order contribution to the Feynman-Vernon action. While this is the standard term having the same functional form also in the case of bosonic harmonic baths, the dependence on temperature will in general be different for a fermionic bath with two-fermion coupling.
We identified one regime where nevertheless the fermionic environment mimics a bosonic one.                 
Finally, we showed (details in appendix) that the fourth order terms in the generalized Feynman-Vernon influence action vanish for the model considered. The first non-zero corrections to Feynman-Vernon or Keldysh theory are hence of sixth order in the system-bath interaction coefficient.

\section*{Acknowledgement}
\label{ack}
This work was supported  by  the  Foundation for Polish Science through TEAM-NET project (contract no.POIR.04.04.00-00-17C1/18-00)  (EA)
and by the European Research Council under grant 742104 (JT).

\bibliographystyle{unsrt}
\bibliography{refs,quantum-fluctuations,quantum-computing,lit,polaron}

\appendix

\section{Non-harmonic baths and cluster expansions}
\label{sec:non-harmonic}
Here we briefly sketch how the cumulant expansion can be used to express influence of the bath on the system. Firstly we summarize necessary notation from~\cite{AurellGoyalKawai2019}. That paper employs the super-operator approach to find dynamics of the system interacting with the environment: A map governing evolution of system operators is obtained by tracing out bath degrees of freedom from the formal solution of the full (i.e. including system and the bath)  Liouville–von Neumann equation. A crucial step in performing the trace is evaluation of multi-time super-operator correlation functions (correlation functions with indices) 
in the bath, which are defined in terms of ordinary bath correlation
by
\begin{eqnarray}\label{eq:multi-time-correlation-reorder}
   C^{d_1,\cdots,d_n}(t_1,\cdots,t_n) &=&  \tr\big[
\overrightarrow{\mathcal{T}}\sb \left (\prod_{d_i = "<"} \mathcal{Q}^{d_i}\sb (t_i) \right ) \\ \nonumber
&& \qquad \overleftarrow{\mathcal{T}}\sb \left (\prod_{d_i = ">"} \mathcal{Q}^{d_i}\sb (t_i) \right ) \rho_B(t_0) \big],
\end{eqnarray}
where $\mathcal{Q}\sb (t_i)$ are time evolved bath operators from the interaction part of the Hamiltonian (in interaction picture) and $\rho_B(t_0) $ is the initial state of the bath. As the starting point was  Liouville–von Neumann equation, one needs to include indecies $d_1,\cdots,d_n$ to time-order the operators in two groups,
one $(d_i = "<")$ acting from the right on the bath density matrix in ascending time order, and other other ($d_i = ">"$) acting from the left in descending time order.

Let us recall that, for the ordinary operator correlation functions, successive orders of cumulants (cluster expansion)
are defined inductively as
\begin{eqnarray}\label{eq:cumulants-123}
   G_1(t_1) &=&    C(t_1) \nonumber \\
   G_2(t_1,t_2) &=&    C(t_1,t_2) -  G_1(t_1) G_1(t_2)  \nonumber \\ 
   G_3(t_1,t_2,t_3) &=&    C(t_1,t_2,t_3) -   G_1(t_1) G_1(t_2) G_1(t_3) - G_1(t_1) G_2(t_2,t_3)    \nonumber \\
                              &&\quad - G_1(t_2) G_2(t_1,t_3) -  G_1(t_3) G_2(t_1,t_2)     \nonumber \\ 
                              &\vdots&   \nonumber
\end{eqnarray}
The only difference between standard and super-operator correlation functions is that the latter need to be time ordered of time  as determined by the indices $d_1,\ldots,d_{N}$. Once this is done one can write a general cumulant expansion as
\begin{widetext}
\begin{eqnarray}
    C^{d_1, \cdots d_{N}}(t_1, \cdots, t_{N}) &=& \sum_{\text{(all possible groupings)}} \prod_{\text{(groups of one time)}} G_1 (t)
\prod_{\text{(groups of two times)}} G_2 (t,t') \cdots
\label{eq:cumulant-expansion}
\end{eqnarray}
where $N$ can be even or odd, and where the times on the right-hand side are inserted after the re-ordering.
All odd order cumulants vanish for a bath where the Hamiltonian is an even function (as in our case)
and the second order cumulant is the same as the second order correlation function. 
The first non-trivial cumulant is then
\begin{eqnarray}\label{eq:cumulants-4}
   G_4^{d_1,d_2,d_3,d_4}(t_1,t_2,t_3,t_4) &=&    C^{d_1,d_2,d_3,d_4}(t_1,t_2,t_3,d_4) - C^{d_1,d_2}(t_1,t_2) C^{d_3,d_4}(t_3,t_4) 
- C^{d_1,d_3}(t_1,t_3) C^{d_2,d_4}(t_2,t_4) \nonumber \\
&&\quad - C^{d_1,d_4}(t_1,t_4) C^{d_2,d_3}(t_2,t_3) 
\end{eqnarray}
where we have retained the super-operator notation on the right-hand side.
\end{widetext}
All cumulants beyond $G_2$ vanish for correlation functions of (classical) Gaussian processes~\cite{VANKAMPEN}.
This also holds as for operator correlation functions of harmonic bosonic baths,
because in the path integral language these are all determined by Gaussian integrals.
Alternatively, all higher-order operator correlation functions are in a bath of free bosons
by Wick theorem given by combinations of pairwise operator correlation functions, which give
same expressions as the cumulants used here.
For free fermions all higher-order correlation functions are also given in terms of pair-wise
combinations of pair-wise correlation functions, but with signs, and therefore different from
the cumulants used here.

\section{Generalized Feynman-Vernon actions}
\label{sec:generalized-FV}

This Appendix summarizes the derivation of the generalized Feynman-Vernon action
from~\cite{AurellGoyalKawai2019} and relates it to the cluster expansion.
The multi-time super-operator function in
\eqref{eq:multi-time-correlation-reorder} multiplies super-operator representation of the system operator. The connection to the path integral formulation is established in the following way: For indices $d_i="<"$ the super-operators correspond  to forward paths $X_i$, whereas for indices $d_i=">"$ to a backward paths $Y_i$ (with a negative sign). 
A general bath correlation function is represented as in Eq. (\ref{eq:cumulant-expansion}) then a relevant series summation is performed. As a result, one obtains a reduced system propagator of the form (\ref{eq:quantum-map-dissipative-2}), where contributions to the 
generalized Feynman-Vernon action $S^{(n,m)}$ contain $n$ number of $X$ and $m$ number $Y$ as
\begin{widetext}
\begin{eqnarray}\label{eq:cumulants-5}
S^{(n,m)}&=& \frac{(-i)^{n} (i)^{m}}{\hbar^{n+m}}\int_{t_0}^{t}\dd{s_1} \phi(s_1) \int_{t_0}^{s_1}  \dd{s_2}\phi(s_2)\cdots \int_{t_0}^{t}\dd{u_1}\phi(u_1) \int_{t_0}^{u_1}  \dd{u_2}\phi(u_2)\cdots
X_{s_1}X_{s_2}\cdots X_{s_n} \nonumber \\
&&\quad \cdot Y_{u_1} Y_{u_2}\cdots Y_{u_m}  G_{n+m}(u_m,\ldots,u_1,s_1,\ldots,s_n).
\end{eqnarray}
The last term in the above expression is the cumulant of the operator correlation function with an appropriate time ordering (first times for the backward path
in reverse chronological order, then times for the forward path in chronological order). The term corresponding to $m+n=2$ is the standard quadratic Feynman-Vernon action as given by Eq. (\ref{eq:quadraticFV}).

Renaming the variables so that times are always ordered $s>u$ and rewriting the resulting expression in terms of sum and difference of system paths  $\chi_s=X_s+Y_s$ and $\xi_s=X_s-Y_s$  gives
\begin{eqnarray}
\label{eq:second-cumulant}
\sum_{n+m=2}S^{(n,m)} &=& -\frac{1}{2} \int_{t_0}^{t} \dd{s} \xi_s \phi(s) \int_{t_0}^{s} \dd{u}\phi(u) \left(\chi_u A +\xi_u B \right)
\end{eqnarray}
where $A,B$ are difference and sum of bath correlation functions at different times 
\begin{eqnarray}
\label{eq:second-cumulant-2}
A &=& C(s,u) - C(u,s) \\
B &=& C(s,u) +C(u,s).
\end{eqnarray}
We want to simplify the above expression with regard to the correlation function and shift all time re-orderings to the system operators.  Therefore we rewrite it as
\begin{eqnarray}
\label{eq:Adquadratic}
\int^{t}_{t_i} dt_1 \int^{t}_{t_i} dt_2 C(t_1,t_2) \left[ \Theta(t_2-t_1) X_{t_1} X_{t_2} + \Theta(t_1-t_2) Y_{t_1} Y_{t_2} - Y_{t_1} X_{t_2}  \right],
\end{eqnarray}
where $\Theta(t-t')$ is the Heaviside step function.
The next term of the action is a sum of all contributions of total order three, however in our case it vanishes and will be not discussed here. 

\section{Calculation of the 4th order cumulant}
\label{app:4th}
In order to show that in the considered model the fourth-order cumulant vanishes we will exploit several properties of the super-operator expression for the fourth order Feynman-Vernon action, which reads
\begin{equation}
\label{eq:operatorFV4th}
\frac{(-i)^{4}}{4 !\hbar^4} \int_{t_{i}}^{t_{1}} \mathrm{d} t_{1} \int_{t_{i}}^{t_{f}} \mathrm{d} t_{2} \int_{t_{i}}^{t_{f}} \mathrm{d} t_{3}
\int_{t_{i}}^{t_{f}} \mathrm{d} t_{4}
 \sum_{d_{1}, d_{2}, d_{3},d_{4}} G_{4}^{d_{1}, d_{2} ; d_{3}, d_{4}}\left(t_{1}, t_{2}, t_{3}, t_4\right) \mathcal{X}_{\mathrm{s}}^{d_{1}}\left(t_{1}\right) \mathcal{X}_{\mathrm{s}}^{d_{2}}\left(t_{2}\right) \mathcal{X}_{\mathrm{s}}^{d_{3}}\left(t_{3}\right) \mathcal{X}_{\mathrm{s}}^{d_{4}}\left(t_{4}\right),
\end{equation}
where
\begin{equation}
\begin{aligned}
G_{4}^{d_{1}, d_{2}, d_{3}, d_{4}}\left(t_{1}, t_{2}, t_{3}, t_{4}\right) &=C^{d_{1}, d_{2}, d_{3}, d_{4}}\left(t_{1}, t_{2}, t_{3}, d_{4}\right)-C^{d_{1}, d_{2}}\left(t_{1}, t_{2}\right) C^{d_{3}, d_{4}}\left(t_{3}, t_{4}\right) \\
&-C^{d_{1}, d_{3}}\left(t_{1}, t_{3}\right) C^{d_{2}, d_{4}}\left(t_{2}, t_{4}\right)-C^{d_{1}, d_{4}}\left(t_{1}, t_{4}\right) C^{d_{2}, d_{3}}\left(t_{2}, t_{3}\right),
\end{aligned}
\end{equation} 
 is the fourth order super-operator cumulant involving super-operator correlation functions of the bath
 \begin{eqnarray}
 \begin{aligned}
 C^{d_{1}, \cdots, d_{n}}\left(t_{1}, \cdots, t_{n}\right) &= \sum_{k_1, \ldots k_{2n}, l_1, \ldots, k_{2n}} g_{k_1 l_1} \ldots g_{k_{2n} l_{2n}}  Tr\left[\overleftarrow{\mathcal{T}} (\hat{Q}_{k_1}(t_1)\hat{Q}_{l_1}(t_1) \dots \hat{Q}_{k_{2n}}(t_{2n})\hat{Q}_{l_2n}(t_{2n}) ) \rho\right].
 \end{aligned}
 \end{eqnarray}
 As for now $d_{1}, d_{2}, d_{3}, d_{4}$ will be fixed and dependence on them will be dropped. The key step in providing expression for the cumulant is calculation of the fourth order correlation function, which for our model reads
\begin{eqnarray}
\label{eq:4timecorfun}
 C(t_1,t_2,t_3,t_4) = \sum_{k_1,l_1,k_2,l_2,k_3,l_3,k_4,l_4}g_{k_1l_1}g_{k_2l_2}g_{k_3l_3}g_{k_4l_4}Tr \left[\hat{Q}_{k_1}(t_1)\hat{Q}_{l_1}(t_1)\hat{Q}_{k_2}(t_2)\hat{Q}_{l_2}(t_2)\hat{Q}_{k_3}(t_3)\hat{Q}_{l_3}(t_3)\hat{Q}_{k_4}(t_4)\hat{Q}_{l_4}(t_4) \rho \right]. \nonumber \\
\end{eqnarray}
In the above there will be four fermionic operators with indices $k_i$ as well as four with indices $l_j$. The non-zero contributions to the action come from pairing of indices: Expressions with an odd number of an indices $k_i, l_i$ vanish. In order to provide the final result in the simplest form we deal with different possible pairings of the fermion operators case by case. We also note that the time integrals over $t_1, t_2, t_3, t_4$ in Eq. (\ref{eq:operatorFV4th}) are unconstrained, although they can also be written as time-ordered integrals. In order to avoid confusion with the notation time-ordered times will be denoted as $s_1, s_2, s_3, s_4$.
\subsection{Case I: Pair-wise groupings}
\label{sub:subI}
Here we treat the terms, in which if $k_i=k_j$ or $k_i=l_j$ from some indices $i$ and $j$, then they are different from all the other $k$ and $l$ (e.g. $ k_1=k_2=k \; k_3=k_4=k'$). Furthermore, we can divide those terms into two categories: "Separated-pairing" and "mixed pairing" terms. For "Separated-pairing"  $k$ indices are paired among each other and the same holds for $l$. In "mixed pairing" terms $k$'s might br paired with $l$'s. Subsequently, we observe that "mixed-pairing" terms can be brought into "separated-pairing" form with the help of a permutation $k_j \leftrightarrow l_j$ (note that such permutations do not change the overall sign of terms). As a result, is is sufficient to consider "separated-pairing" terms with an additional factor $4$ and in such a way all the terms are accounted for. Now we use the fact that all fermionic operators with $k$ indices can be moved to the left and, if time ordering is preserved, the sign of this expression remains the same. The possible pairings for $k$'s are 
\begin{eqnarray}
&&\hat{Q}_k(s_1)\hat{Q}_k(s_2) \hat{Q}_{k'}(s_3) \hat{Q}_{k'}(s_4) \equiv [12]_{k}[34]_{k'} \\
&&\hat{Q}_k(s_1)\hat{Q}_{k'}(s_2) \hat{Q}_{k}(s_3) \hat{Q}_{k'}(s_4) = - \hat{Q}_k(s_1)\hat{Q}_{k}(s_3)  \hat{Q}_{k'}(s_2)  \hat{Q}_{k'}(s_4) \equiv - [13]_{k}[24]_{k'} \\   
&&\hat{Q}_k(s_1)\hat{Q}_{k'}(s_2) \hat{Q}_{k'}(s_3) \hat{Q}_{k}(s_4) =  \hat{Q}_k(s_1) \hat{Q}_{k}(s_4) \hat{Q}_{k'}(s_2) \hat{Q}_{k'}(s_3)   \equiv [14]_{k}[23]_{k'},
\end{eqnarray}
where we introduced a symbolic notation: The first and second square bracket groups operators with $k$ and $k'$ respectively and number inside bracket  denote indices of re-ordered times $s_i$. To get the total operator acting on $\rho_B$ in Eq. (\ref{eq:4timecorfun}) one needs to multiply the $k,k;$ operators with $l, l'$ operators and include appropriate combination of coupling constants. In fact there are just two possible pre-factors: $g_{kl}^2 g_{k'l'}^2$ for terms of a form $[ab]_{k}[cd]_{k'}[ab]_{l}[cd]_{l'}$ and  $g_{kl} g_{kl'} g_{k'l} g_{k'l'}$ for all others. It will prove convenient to write the resulting expression in the following form 
\begin{eqnarray}
&& \left(g_{kl}^2 g_{k'l'}^2 -g_{kl} g_{kl'} g_{k'l} g_{k'l'}\right)  \left( [12]_{k}[34]_{k'}  [12]_{l}[34]_{l'} + [13]_{k}[24]_{k'} [13]_{l}[24]_{l'} + [14]_{k}[23]_{k'} [14]_{l}[23]_{l'} \right) + \label{eq:same-same} \\
&&\left(g_{kl}^2 g_{k'l'}^2 + g_{kl} g_{kl'} g_{k'l} g_{k'l'}  \right)\left([12]_{k}[34]_{k'}-[13]_{k}[24]_{k'}+[14]_{k}[23]_{k'}\right)\left([12]_{l}[34]_{l'}-[13]_{l}[24]_{l'}+[14]_{l}[23]_{l'}\right), \label{eq:all-all}
\end{eqnarray}
The above expressions are written using re-ordered times $s_1,s_2,s_3,s_4$. Assuming that superoperator indices $d_1,d_2,d_3,d_4$ are fixed we relate re-ordered times to unconstrained times in the following way
\begin{equation}
s_1 = t_x \;\;\; s_2 = t_y \;\;\; s_3 = t_z\;\;\;s_4 = t_w,
\end{equation}
where indices ${x,y,z,w}$  are related to values in${1,2,3,4}$ through a permutation
\begin{equation}
(x,y,z,w)=P_{d,t}(1,2,3,4),
\end{equation} 
where indices $d,t$ indicate dependence of the permutation on the superoperator ordering $d_i$ and unconstrained times $t_j$. Then we have that e.g. $[12]_{k}= \hat{Q}_k(s_1) \hat{Q}_k(s_2)= \hat{Q}_k(t_a) \hat{Q}_k(t_b)$. We now show that Eq. (\ref{eq:all-all})  vanish.  Consider a new $d$-dependent time ordering  of operators $\hat{Q}_m(t)$, where $m \in \left\{k,k',l,l \right\}$
\begin{eqnarray}
&&S_{d,m}(x,y) = T_{d,m}(x,y) = \overrightarrow{\mathcal{T}} \prod_{i\in (x,y); d_i=-} \hat{Q}_m(t_i) \overleftarrow{\mathcal{T}}\prod_{j\in (x,y); d_i=+} \hat{Q}_m(t_j),
\end{eqnarray}
if the order of the operators is the same as $(x,y)$, and
\begin{eqnarray}
&&S_{d,m}(x,y) = - T_{d,m}(x,y),  
\end{eqnarray}
if the order of the operators is the opposite of $(x,y)$.
Additionally we will need the sign of a permutation $S(m \rightarrow n)$, then we can rewrite the first bracket in Eq. (\ref{eq:all-all}) as
\begin{eqnarray}
&&[12]_{k}[34]_{k'}-[13]_{k}[24]_{k'}+[14]_{k}[23]_{k'} =  S(1234 \rightarrow xyzw) S_{d,k}(a,b) S_{d,k'}(c,d) + S(1234 \rightarrow xzyw) S_{d,k}(a,c) S_{d,k'}(b,d) +  \nonumber \\
 &&S(1234 \rightarrow xwyz) S_{d,k}(a,d) S_{d,k'}(b,c).
\end{eqnarray}
The above expression is a  sum that goes over 3 permutations of $(xywz)$ where the first is the identity, $S(abcd \rightarrow abcd) = 1$. It can be extended to the sum over all 24 permutations of $(abcd)$ 
\begin{eqnarray}
[12]_{k}[34]_{k'}-[13]_{k}[24]_{k'}+[14]_{k}[23]_{k'} = \frac{1}{8} \left[  \sum_{xyzw} S(1234 \rightarrow  xyzw) S_{d,k}(x,y) S_{d,k'}(z,w) \right].
\end{eqnarray}
The same argument applies to the $l,l'$ terms in the second bracket of Eq. (\ref{eq:all-all}) so we can rewrite both brackets as 
\begin{eqnarray}
\frac{1}{64} \left[  \sum_{xyzw} S(1234 \rightarrow  xyzw) S_{d,k}(x,y)S_{d,k'}(z,w) \right]\left[  \sum_{x'y'z'w'} S(1234 \rightarrow  x'y'z'w') S_{d,k}(x',y') S_{d,k'}(z',w') \right].
\end{eqnarray}
Let us consider product of two terms from the two sums
\begin{eqnarray}
\frac{1}{64}  S(1234 \rightarrow  xyzw) S_{d,k}(x,y)  S_{d,k'}(z,w)S(1234 \rightarrow  x'y'z'w') S_{d,k}(x',y')S_{d,k'}(z',w').
\end{eqnarray}
We need to consider the following cases: \\
1. The same pairing i.e.
\begin{itemize}
	\item $xy=x'y'$ and $zw=z'w'$
	\item $xy=z'w'$ and $zw=x'y'$
\end{itemize}
Those terms are of the same structure as the ones in Eq. (\ref{eq:same-same}).  \\
2. Different pairing e.g. $\frac{1}{64} S(1234 \rightarrow xyzw) S_{d,k}(x,y) S_{d,k'}(z,w) S(1234 \rightarrow xzyw) S_{d,l}(x,z) S_{d,l'}(y,w)$. The overall expression is summed over indices $d_i$ and integrated over times $t_j$. Consider therefore the following change of variables 
\begin{eqnarray}
t'_x=t_y \; \; \; d'_x = d_y \; \; \;  t'_y=t_x \; \; \;  d'_y = d_x,
\end{eqnarray}
with rest of them unchanged. Now we will analyze how such a change affects the sign of the considered term (it is useful to bring in the dependence of $t$ and $t'$).  We have 
\begin{eqnarray}
&&S(1234 \rightarrow xyzw) \;\;\; \text{unchanged}, \nonumber \\ 
&&S_{d,k}(x,y;t) \rightarrow S_{d',k}(x,y;t')  \;\;\; \text{unchanged}, \nonumber \\ 
&&S_{d,k}(z,w;t) \rightarrow S_{d',k}(z,w;t')  \;\;\; \text{unchanged}, \nonumber \\ 
&&S(1234 \rightarrow xzyw) \;\;\; \text{unchanged}, \nonumber \\ 
&&S_{d,k}(x,z;t) \rightarrow S_{d',k}(x,z;t') \;\;\; \text{changed}, \nonumber \\ 
&&S_{d,k}(y,w;t) = S_{d',k'}(y,w;t') \;\;\; \text{changed}. \nonumber
\end{eqnarray}
We compare  the above to the effect of permuting indices $x \leftrightarrow y$. 
\begin{eqnarray}
&&S(1234 \rightarrow xyzw) \rightarrow S(1234 \rightarrow yxzw) = - S(1234 \rightarrow xyzw),   \nonumber \\ 
&&S_{d,k}(x,y;t) \rightarrow S_{d,k}(y,x;t) = - S_{d,k}(x,y;t),   \\ 
&&S_{d,k}(z,w;t) \;\;\; \text{unchanged}, \nonumber \\ 
&&S(1234 \rightarrow xzyw)  \rightarrow S(1234 \rightarrow yzxw)  =-S(1234 \rightarrow xzyw)  \;\;\; \text{unchanged}, \nonumber \\ 
&&S_{d,k}(x,z;t) \rightarrow S_{d,k}(y,z;t) \;\;\; \text{changed}, \nonumber \\ 
&&S_{d,k}(y,w;t) = S_{d,k'}(x,w;t) \;\;\; \text{changed}. \nonumber
\end{eqnarray}
 From definition of $S_{d,k}(x,z;t)$ it follows that 
\begin{eqnarray}
S_{d',k}(x,z;t') = S_{d,k}(y,z;t) \;\;\;  \text{and} \;\;\;  S_{d',k}(y,w;t')  =  S_{d,k}(x,w;t),
\end{eqnarray}
and all the above same relations hold for $k',l,l'$. Therefore we found that all terms where the pairing is not the same cancel pairwise. 
As a result the only non-zero term, from Eq. (\ref{eq:same-same}) and (\ref{eq:all-all}) is 
\begin{equation}
 g_{kl}^2 g_{k'l'}^2  \left( [12]_{k}[34]_{k'}  [12]_{l}[34]_{l'} + [13]_{k}[24]_{k'} [13]_{l}[24]_{l'} + [14]_{k}[23]_{k'} [14]_{l}[23]_{l'} \right).
\end{equation}
Performing the trace yields the following result 
\begin{equation}
\label{eq:counterIres}
\sum_{k,l,k',l'}g_{kl}^2 g_{k'l'}^2  \left( \left \langle 12 \right \rangle_{k} \left \langle 34 \right \rangle_{k'}   \left \langle 12 \right \rangle_{l} \left \langle 34 \right \rangle_{l'} + \left \langle 13 \right \rangle_{k} \left \langle 24 \right \rangle_{k'}   \left \langle 13 \right \rangle_{l} \left \langle 24 \right \rangle_{l'} + \left \langle 14 \right \rangle_{k} \left \langle 23 \right \rangle_{k'}   \left \langle 14 \right \rangle_{l} \left \langle 23 \right \rangle_{l'} \right),
\end{equation}
where $\left \langle ab \right \rangle_{k} \left \langle ab \right \rangle_{m}\equiv \left \langle \hat{Q}_m(t_a) \hat{Q}_m(t_b) \right \rangle $ is thermal expectation value and we restored the sum over bath degrees of freedom.
\subsection{Case II: Four $k$ grouping, pairwise $l$ grouping }
\label{sub:subII}
An example of such indices arrangement is $k_1=k_2=k_3=k_4=k \; l_1=l_2=l \; l_3=l_4=l'$. Using the operator notation introduced in the previous case we find that the relevant expression is
\begin{equation}
g_{kl}^2 g_{kl'}^2[1234]_k \left([12]_{l}[34]_{l'}-[13]_{l}[24]_{l'}+[14]_{l}[23]_{l'}\right)
\end{equation}
This can be evaluated into
\begin{eqnarray}
g_{kl}^2 g_{kl'}^2\left(\left \langle 12 \right \rangle_{k} \left \langle 34 \right \rangle_{k} -\left \langle 13 \right \rangle_{k} \left \langle 24 \right \rangle_{k} +\left \langle 14 \right \rangle_{k} \left \langle 23 \right \rangle_{k} \right)\left(\left \langle 12 \right \rangle_{k} \left \langle 34 \right \rangle_{l'} -\left \langle 13 \right \rangle_{k} \left \langle 24 \right \rangle_{l'} +\left \langle 14 \right \rangle_{k} \left \langle 23 \right \rangle_{l'} \right),
\end{eqnarray}
where $\left \langle ab \right \rangle_{k}$ is thermal expectation value. The above expression can be simplified using exactly the same discussion as in the previous Subsection. The reason for this is that it does not involve indices $k$'s and $l$'s but only time orderings and signs of permutations. Therefore we find that the final expression is
\begin{eqnarray}
\label{eq:caseIIres}
\sum_{k,l,l'}g_{kl}^2 g_{kl'}^2\left(\left \langle 12 \right \rangle_{k} \left \langle 34 \right \rangle_{k} \left \langle 12 \right \rangle_{l} \left \langle 34 \right \rangle_{l'}  +\left \langle 13 \right \rangle_{k} \left \langle 24 \right \rangle_{k} \left \langle 13 \right \rangle_{l} \left \langle 24 \right \rangle_{l'} +\left \langle 14 \right \rangle_{k} \left \langle 23 \right \rangle_{k} \left \langle 14 \right \rangle_{l} \left \langle 23 \right \rangle_{l'} \right),
\end{eqnarray}
where we restored the sum over bath degrees of freedom.

\subsection{Case III: Four $k,l$ grouping}
\label{sub:subIII}
In the operator notation we can write
\begin{eqnarray}
g^4_{kl}[1234]_k [1234]_l,
\end{eqnarray}
using Wicks theorem this evaluates into
\begin{equation}
g^4_{kl}\left( \left \langle 12 \right \rangle_{k} \left \langle 34 \right \rangle_{k} -\left \langle 13 \right \rangle_{k} \left \langle 24 \right \rangle_{k} +\left \langle 14 \right \rangle_{k} \left \langle 23 \right \rangle_{k}   \right)\left( \left \langle 12 \right \rangle_{l} \left \langle 34 \right \rangle_{l} -\left \langle 13 \right \rangle_{l} \left \langle 24 \right \rangle_{l} +\left \langle 14 \right \rangle_{l} \left \langle 23 \right \rangle_{l}   \right). 
\end{equation}
Here again we can apply reasoning from Subsection (\ref{sub:subI}), so finally we have
\begin{eqnarray}
\label{eq:caseIIIres}
\sum_{k,l}g_{kl}^4\left(\left \langle 12 \right \rangle_{k} \left \langle 34 \right \rangle_{k} \left \langle 12 \right \rangle_{l} \left \langle 34 \right \rangle_{l}  +\left \langle 13 \right \rangle_{k} \left \langle 24 \right \rangle_{k} \left \langle 13 \right \rangle_{l} \left \langle 24 \right \rangle_{l} +\left \langle 14 \right \rangle_{k} \left \langle 23 \right \rangle_{k} \left \langle 14 \right \rangle_{l} \left \langle 23 \right \rangle_{l} \right),
\end{eqnarray}
\subsection{Counter terms and the final result}
Subsections \ref{sub:subI}, \ref{sub:subII}, \ref{sub:subIII} were devoted to calculation of correlation function involving four times. In order to obtain final expression for the fourth order cumulant one needs to subtract from those results products of two times correlation functions. We find 
\begin{eqnarray}
&&C(t_1,t_2)C(t_3,t_4) + C(t_1,t_3)C(t_2,t_4) + C(t_1,t_4)C(t_2,t_3) = \nonumber \\ \label{eq:counterI}
&&\sum_{k,k',l,l'} \left( \left \langle 12 \right \rangle_{k} \left \langle 34 \right \rangle_{k'}   \left \langle 12 \right \rangle_{l} \left \langle 34 \right \rangle_{l'} + \left \langle 13 \right \rangle_{k} \left \langle 24 \right \rangle_{k'}   \left \langle 13 \right \rangle_{l} \left \langle 24 \right \rangle_{l'} + \left \langle 14 \right \rangle_{k} \left \langle 23 \right \rangle_{k'}   \left \langle 14 \right \rangle_{l} \left \langle 23 \right \rangle_{l'} \right) \\ \label{eq:counterII}
&&\sum_{k,l,l'} \left(\left \langle 12 \right \rangle_{k} \left \langle 34 \right \rangle_{k} \left \langle 12 \right \rangle_{l} \left \langle 34 \right \rangle_{l'}  +\left \langle 13 \right \rangle_{k} \left \langle 24 \right \rangle_{k} \left \langle 13 \right \rangle_{l} \left \langle 24 \right \rangle_{l'} +\left \langle 14 \right \rangle_{k} \left \langle 23 \right \rangle_{k} \left \langle 14 \right \rangle_{l} \left \langle 23 \right \rangle_{l'} \right) \\ \label{eq:counterIII}
&&\sum_{k,l,} g_{kl}^42\left(\left \langle 12 \right \rangle_{k} \left \langle 34 \right \rangle_{k} \left \langle 12 \right \rangle_{l} \left \langle 34 \right \rangle_{l}  +\left \langle 13 \right \rangle_{k} \left \langle 24 \right \rangle_{k} \left \langle 13 \right \rangle_{l} \left \langle 24 \right \rangle_{l} +\left \langle 14 \right \rangle_{k} \left \langle 23 \right \rangle_{k} \left \langle 14 \right \rangle_{l} \left \langle 23 \right \rangle_{l} \right).
\end{eqnarray}
 Compering the above to the results of Subsections \ref{sub:subI}, \ref{sub:subII}, \ref{sub:subIII} we find that line (\ref{eq:counterI}) equals to Eq. (\ref{eq:counterIres}), line (\ref{eq:counterII}) equals to Eq. (\ref{eq:caseIIres}) and line (\ref{eq:counterIII}) equals to Eq. (\ref{eq:caseIIIres}). As a result, in our model the fourth order cumulant vanishes.  
\end{widetext}

\end{document}